\documentclass{gji}
\usepackage{timet}
\usepackage{graphicx}

\title[The SGR 1806-20 magnetar signature on the Earth's magnetic field]
  {The SGR 1806-20 magnetar signature on the Earth's magnetic field}
\author[M. Mandea and G. Balasis]
{M. Mandea$^1$ and G. Balasis$^2$ \\
$^1$GeoForschungsZentrum Potsdam, Telegrafenberg, \emph{14473} Potsdam, Germany \\
$^2$Institute for Space Applications and Remote Sensing, National Observatory of Athens, 
Metaxa and Vas. Pavlou, Palea Penteli, \emph{15236} Athens, Greece}

\date{Received ; in original form }
\pagerange{\pageref{firstpage}--\pageref{lastpage}}
\volume{}
\pubyear{}

\begin{document}

\label{firstpage}

\maketitle

\begin{summary}
SGRs denote ``soft $\gamma$-ray repeaters'', a small class of slowly spinning 
neutron stars with strong magnetic fields. On 27 December 2004, a giant flare 
was detected from magnetar SGR 1806-20. The initial spike was followed by a 
hard-X-ray tail persisting for 380 s with a modulation period of 7.56 s. This 
event has received considerable attention, particularly in the astrophysics 
area. Its relevance to the geophysics community lies in the importance of 
investigating the effects of such an event on the near-earth electromagnetic 
environment. However, the signature of a magnetar flare on the geomagnetic 
field has not previously been investigated. Here, by applying wavelet 
analysis to the high-resolution magnetic data provided by the CHAMP 
satellite, a modulated signal with a period of 7.5 s over the duration of the 
giant flare appears in the observed data. Moreover, this event was detected 
by the energetic ion counters onboard the DEMETER satellite.
\end{summary}

\begin{keywords}
geomagnetic field, magnetar, wavelet analysis
\end{keywords}

\section{Introduction}

SGRs are galactic X-ray stars that emit, during sporadic times of high 
activity, a large number of short-duration (around 0.1 s) bursts of hard 
X-rays (Duncan and Thompson, 1992). A SGR is thought to be a magnetar, being 
a strongly magnetized neutron star powered by a very strong magnetic field 
($\ge$ 10$^{15}$ Gauss). On 27 December 2004 a powerful burst of X- and 
$\gamma$-rays from one of the most highly magnetized neutron stars 
(SGR 1806-20) of our Galaxy reached the Earth's environment (Hurley et al., 
2005). The Solar system received a shock, which is thought to be due to a 
cataclysm in the magnetar that caused it to emit as much energy in two-tenths 
of a second as the Sun gives off in 250,000 years. The signature of this 
event on the Earth's magnetic field has not previously been investigated. 
Here, we present the first results of the magnetar footprints on magnetic 
data recorded by near-Earth satellites. The magnetar SGR 1806-20 is the third 
such event ever recorded along with two others that were noted in 1979 and 
1998 (Mazets et al., 1979; Hurley et al., 1999).

Several properties of this magnetar flare are relevant to our study. Firstly, 
a precursor of $\sim$ 1 s was observed 142 s before the flare, with a roughly 
flat-topped profile (Hurley et al., 2005). The intensity of the main initial 
spike saturated all X- and $\gamma$-ray detectors. However, particle 
detectors on board of RHESSI and Wind spacecraft (Boggs et al., 2004; Mazets 
et al., 2004) were able to record reliable measurements. Several instruments 
designed for other purposes provided important information, as Geotail 
(Terasawa et al., 2005) and Cluster/Double star (Schwartz et al., 2005). The 
first spike was followed by a tail lasting 380 s, during which 7.56 s 
pulsations were clearly observed, by the $\gamma$-ray detectors on board of 
RHESSI (Hurley et al., 2005).

Secondly, a disturbance of the Earth's ionosphere was simultaneously observed 
with the detection of the burst from SGR 1806-20 (Inan et al., 2005). This 
sudden ionospheric disturbance (SID) was recorded as a change in the signal 
strength from very low frequency (VLF) radio transmitters, being noticed by 
stations around the globe (Campbell et al., 2005). These changes in the radio 
signal strength were caused by X-rays arriving from SGR 1806-20, which 
ionized the upper atmosphere and modified the radio propagation properties of 
the Earth's ionosphere (see clearing house of SID data associated with SGR 
1806-20 flare at 
http://www.aavso.org/observing/programs
/solar/sid-sgr1806.shtml). One such 
observation of this ionospheric signature resides within a 21.4 kHz signal 
that originates in Hawaii and propagates along an ionosphere wave guide to 
Palmer Station, Antarctica (Inan et al., 2005). This wave guide is some 
$\sim$ 10,000 km in path length (Inan et al., 2005). As explained above, 
this is not a direct radio detection of SGR 1806-20 (see also 
http://gcn.gsfc.nasa.gov/gcn3/2932.gcn3). Moreover, due to the sub-burst longitude and 
latitude (Inan et al., 2005) and to the geographical distribution of LF/VLF 
beacons and monitoring stations, this burst was not detected by active 
monitoring stations in Germany, Australia, or Canada (Campbell et al., 2005). 
Here, we note that ionospheric disturbances were also reported in the case of 
the magnetar observed in 1998 (Inan et al., 1999). In the case of the 1998
magnetar the flare illuminated the nightside of the Earth and ionized the 
lower ionosphere to levels usually found only during daytime. The magnetar 
responsible for the 2004 burst was about the same distance as the magnetar 
responsible for the 1998 burst, but within 5.25$^\circ$ of the Sun as 
viewed from Earth. Therefore its $\gamma$-rays arrived on the dayside of our 
planet. The 2004 flare 
changed the ionic density at an altitude of 60 km by six orders of magnitude 
(Inan, 2006). It is thus plausible that this change in the ionospheric 
conductivity can cause oscillating perturbations in the current-generated 
magnetic field.

The thrust of this study is to find signatures associated with the explosion 
of the magnetar SGR 1806-20 within satellite measurements of Earth's 
electromagnetic field. Currently, the Earth's electromagnetic field is 
monitored by a number of Low Earth Orbit (LEO) satellite missions. After the 
launch of \O rsted satellite in 1999, the knowledge of the near-Earth 
electromagnetic field has been dramatically improved (Hulot et al., 
2002; L\"uhr et al., 2002; Maus et al., 2002; Tyler et al., 2003; Balasis et 
al, 2004). Since 2000, \O rsted, CHAMP and SAC-C satellites have offered a 
continuous flow of high quality magnetic field measurements. Additionally, 
the DEMETER satellite provides 1 Hz energetic electron detector data. 
Finally, let us note that all these LEO magnetic missions are flying between
the Earth's surface, where the temporal variations of the magnetic field are
continuously monitored by geomagnetic observatories, and the magnetosphere,
where an in-situ investigation of the three-dimensional and
time-varying phenomena is done by the four identical spacecraft of Cluster II
mission.

\section{Data Processing}

We have considered \O rsted (25 Hz scalar data), CHAMP (50 Hz vector data), 
and SAC-C (20 Hz vector data) satellite magnetic data, as well as the 1 Hz 
data from DEMETER satellite energetic electron detector. The giant flare 
occurred during a time span characterized by $k_p$=3- and $D_{st}$=-17 nT, 
i.e., indicating conditions of low geomagnetic activity. All available 
magnetic and electric field data recorded during the giant flare were 
analyzed using wavelet methods (Alexandrescu et al., 1995; Balasis et al., 
2005).

The advantage of analyzing a signal with wavelets is that it enables the 
study of very localized features in the signal (Kumar and Foufoula-Georgiou, 
1997). Owing to its unique time-frequency localization, wavelet analysis is 
especially useful for signals that are non-stationary, have short-lived 
transient components, have features at different scales or have 
singularities, as in the case of the signal due to magnetar SGR 1806-20. The 
basic idea can be understood as a time-frequency plane that indicates the 
frequency content of a signal at every time. The decomposition pattern of the 
time-frequency plane is predetermined by the choice of the basis function. In 
the present study, we used the continuous wavelet transform with the Morlet 
wavelet as the basis function. The results were checked for consistency using 
the Paul and DOG mother functions (Torrence and Compo, 1998).

\section{Results}

In order to find convincing evidence of a causal link between the SGR 1806-20 
flare and the response of the near-earth electromagnetic field, data provided 
by four satellites orbiting the Earth around the time of the flare were 
analyzed. The tracks of these satellites, with direction of flying and
the position of the flare center (Umrad Inan, pers. comm. 2006), are 
indicated in Fig. 1. During the time of the event, both the \O rsted and 
CHAMP satellites were flying over polar regions. The magnetic field 
measurements during the giant flare from the SGR 1806-20 are thus expected to 
be dominated by the polar current systems. This was indeed the case for the 
\O rsted satellite data. 
 
{\bf \O rsted.} This satellite was flying over the South Pole at an altitude
of $\sim$ 700 km. Unfortunately, there are no vector data available over the 
time interval we are interested in. The application of the wavelet transform 
to the scalar data has not shed any light with respect to the 
magnetar signature on the geomagnetic field since these high-frequency 
(25 Hz) data are found to be contaminated both by polar current systems and 
instrumental noise.

{\bf CHAMP.} CHAllenging Minisatellite Payload (CHAMP) is a small near polar, 
low altitude ($\sim$ 430 km) satellite mission. With its highly precise 
magnetometer instruments CHAMP has been generated high quality magnetic field 
measurements for the past 5 years. CHAMP was flying over South Pole region at 
this time, and about 180 s after the SGR 1806-20 outburst, began its 
northward orbital direction. Due to this tracking position, the magnetic 
field measurements are expected to reflect the variations caused by the polar 
magnetospheric-ionospheric current systems. However, the high sampling rate 
data provided by CHAMP satellite offer fine temporal resolution which can be 
exploited by the wavelet analysis in order to detect any signatures caused by 
the magnetar. The wavelet power spectra for the three components of the 
magnetic field around the time of the event are shown in Fig. 2. Strong 
disturbances are observed on a large part of the time interval considered 
here. These fluctuations seem to be random, occurring at almost every 
frequency range in all three components. For this reason, we focus on the 
spectral power for higher frequencies. A high power signal at frequency 
$\sim$ 1 Hz can be associated with the outburst of the magnetar, and is most 
prominent in the East magnetic component. Furthermore, a close up from 21.51 
to 21.524 UT ($\simeq$ 50 s) of the wavelet power spectra is given in Fig. 3. 
Here a finer wavelet resolution is used to better define the period 
corresponding to the maximum signal power. The largest power is clearly 
observed at a period estimation of 7.49 s for North (X) component, 7.46 s for 
East (Y) component and 7.5 s for vertical downward (Z) component, all of 
which are close to the 7.56 s period of modulation of the 380 s long-duration 
signal observed after the initial spike from SGR 1806-20. The differences 
(0.06--0.1 s) between the magnetar period and the period recovered from the 
CHAMP data could be ascribed to the different tools used to determine them. 
Unfortunately, due to strong influences from polar currents, it is hard to 
identify this modulation in CHAMP satellite vector magnetic data for the 
whole duration, and we can only resolve patches of this pulsation (Fig. 2). 
Note, however, that there has been no evidence of pulsation or other signal 
with this frequency in more than 5,000 CHAMP tracks previously investigated 
(Balasis et al., 2005).

{\bf SAC-C.} During the period of this bright flare, the SAC-C satellite was 
flying northwards from the equator, across Africa and Europe at an altitude
of $\sim$ 700 km. This satellite, as Fig. 1 shows, was not in a privileged 
position to detect the SGR 1806-20 event (Campbell et al., 2005). However, 
the wavelet analysis shows an increase in the power spectra at the time of 
the event in $X$ and $Z$ components of the magnetic field (Fig. 4). We note 
that the $Y$ data are too noisy to be considered.

{\bf DEMETER.} This satellite was flying, from north to south over the 
Pacific region during the event. DEMETER (Detection of Electro-Magnetic 
Emissions Transmitted from Earthquake Regions) is a micro-satellite with a 
low altitude ($<$ 800 km) and a nearly polar orbit. The scientific objectives 
of DEMETER are related to the investigation of the ionospheric perturbations 
due to seismic activity, as well as to the global study of the Earth's 
electromagnetic environment (Parrot, 2002). The scientific payload is 
composed of several sensors, the energetic particle analyzer, able to detect 
particles with energy higher than 30 keV, being of a particular interest for 
this study. DEMETER is operated in two modes: a ``Survey'' mode collecting 
averaged data all around the Earth and a ``Burst'' mode collecting high 
sampling data above seismic regions. DEMETER was in an ideal place to detect 
the outburst due to SGR 1806-20 (see the lowermost panel in Fig. 5). The 
particle detector data for three different energy bands (from 90.7 to 526.8 
keV, 526.8 to 971.8 keV and 971.8 to 2342.4 keV) are shown for the orbit, 
including the time of the event (21:30:26 UT). A jump in data at the time of 
the outburst is clearly observed in all panels. Wavelet analysis was also 
performed on these data, but no additional information was derived.

\section{Conclusions}

The effect of the SGR1806-20 flare on the Earth's magnetic field was not 
large, but it was detectable. This first attempt to find a magnetar signature 
in the geomagnetic field clearly indicates that the high resolution CHAMP 
magnetic data are optimal to capture the extremely bright flare from SGR 
1806-20. Indeed, during the first half of the decay phase of the flare a 7.5 
s periodicity is observed in the magnetic field over a magnetically quiet 
period, near the South Pole at 400 km altitude. This observation can be 
explained by a mechanism through which the oscillating flux of ionizing 
$\gamma$-rays could alter the ionospheric conductivity and hence cause 
oscillating perturbations in the current-generated magnetic field.

An attempt to verify this hypothesis for the two previously recorded giant 
flares was not possible since no magnetic satellite missions were operating 
in LEO at that time (i.e., in March 1979 and August 1998). Of course, there 
are many spacecraft carrying magnetometers within the Solar system, but very 
few near planetary ionosphere. For example, the wavelet analysis was
performed on magnetometer data from the Cluster II mission that probes the 
Earth's magnetosphere. In order to be able to visualize, in the wavelet power 
spectrum graph, any significant disturbances of the magnetic field, the power 
spectral density of the signal was amplified by a factor of $2^6$ (in 
comparison to the corresponding spectral density values of the CHAMP data). 
Although there are some indications for a weak pulsation-like signal at 
$\sim$ 8 s, the fact that this signal is almost two orders of magnitudes 
weaker than the one observed in CHAMP data favors the hypothesis of an 
ionospheric origin for the signature found in CHAMP data. 

Furthermore, our analysis can be extended to 1 Hz magnetic data provided by 
ground-based magnetic observatories, but only a small number of them provide 
such high resolution sampling nowadays. Data provided by 12 Canadian
observatories, for which 1 Hz values are available over the period we are
interested in, were also analyzed. For five of these observatories, missing
data or high-level noise, made it difficult to apply the wavelet technique.
For the others, no conclusive evidence for a signature related to SGR 1806-20 
exists.

Analyzing other magnetic data with such a powerful tool as wavelets 
techniques could be relevant for understanding the impact that giant flares 
have on the terrestrial and other planetary magnetic fields. However, the
main difficulty in such studies is due to the availability and quality of
magnetic data. For instance, wavelet analysis of Mars Global Surveyor mission 
magnetic measurements on 27/12/04 was not able to detect any of the magnetar 
features due to the inadequate sampling rate: only 3 s data are now available 
(Michael Purucker, pers. comm. 2005).

\begin{acknowledgments}
The operational support of the CHAMP mission by the German Aerospace Center 
(DLR) is gratefully acknowledged. Constructive suggestions from two anonymous 
reviewers are gratefully acknowledged. We wish to thank Lars 
T\o ffner-Clausen for providing the \O rsted and SAC-C satellite data, 
Michel Parrot for the DEMETER satellite data, Matthias F\"orster, 
Karl-Heinz Fonacon (Braunschweig) for delivering well-calibrated data of
Cluster-FGM and Larry Newitt for the Canadian observatory data.
\end{acknowledgments}

{}

\bsp

\begin{figure*}
\noindent\includegraphics[scale=0.8]{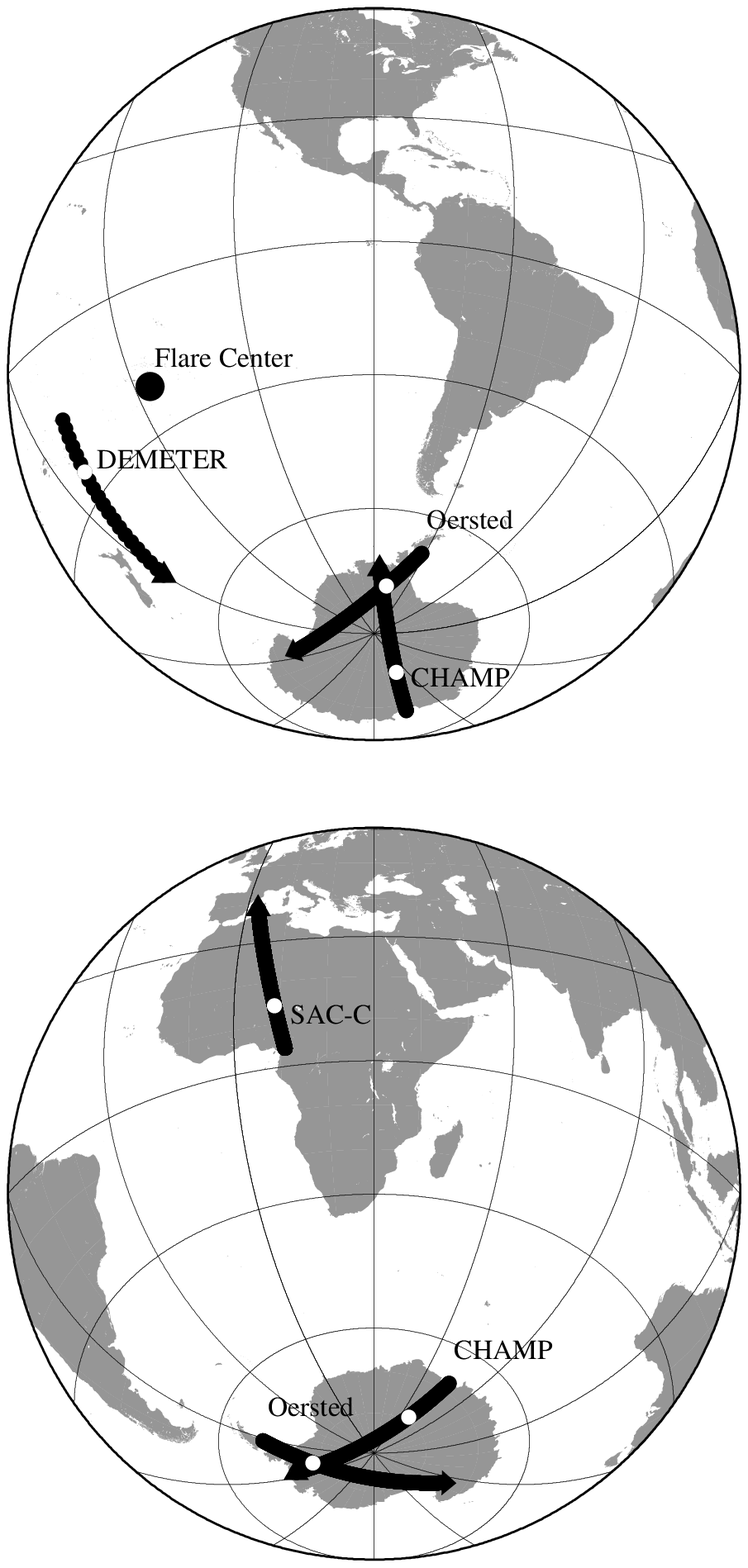}
\vspace{-2cm}
\caption{
The position of CHAMP, \O rsted, SAC-C and DEMETER satellites on 27 December 
2004 (21.46--21.62 UT). The position of each satellite at the time of
the outburst (21:30:26 UT) of the flare from SGR 1806-20 is noted with a 
white circle. The sub-solar point of the source of the flare is also shown.}
\label{fig:1} 
\end{figure*}

\begin{figure*}
\noindent\includegraphics[scale=0.8]{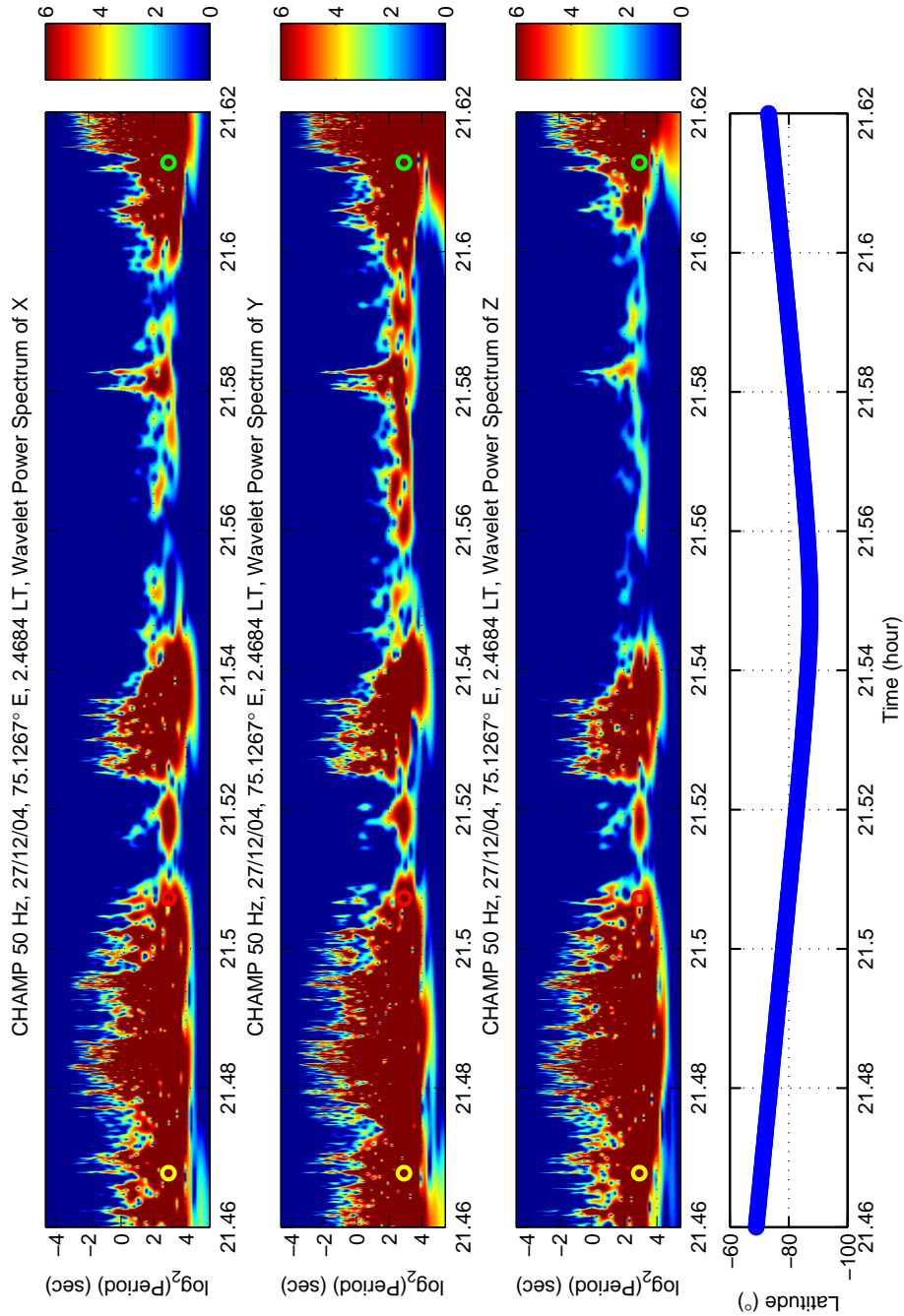}
\caption{
Wavelet analysis of the vector magnetic data provided by the CHAMP 
satellite on 27 December 2004. The main characteristic times of the flare 
from SGR 1806-20, i.e., the precursor (142 s before the flare), outburst, and 
the end of the modulated signal (380 s after the flare), are marked with 
yellow, red and green circles respectively. From top to bottom, wavelet power 
spectra of North (X), East (Y) and vertical downward (Z) components of the
magnetic field are shown. Longitude and local time (LT) are given at the 
beginning of the considered time interval. Strong random fluctuations in all 
three components can be seen at almost all frequency ranges and for the 
largest part of the time interval presented here. The variation of satellite 
latitude with time is given in the fourth panel.}
\label{fig:2} 
\end{figure*}

\begin{figure*}
\noindent\includegraphics[scale=0.8]{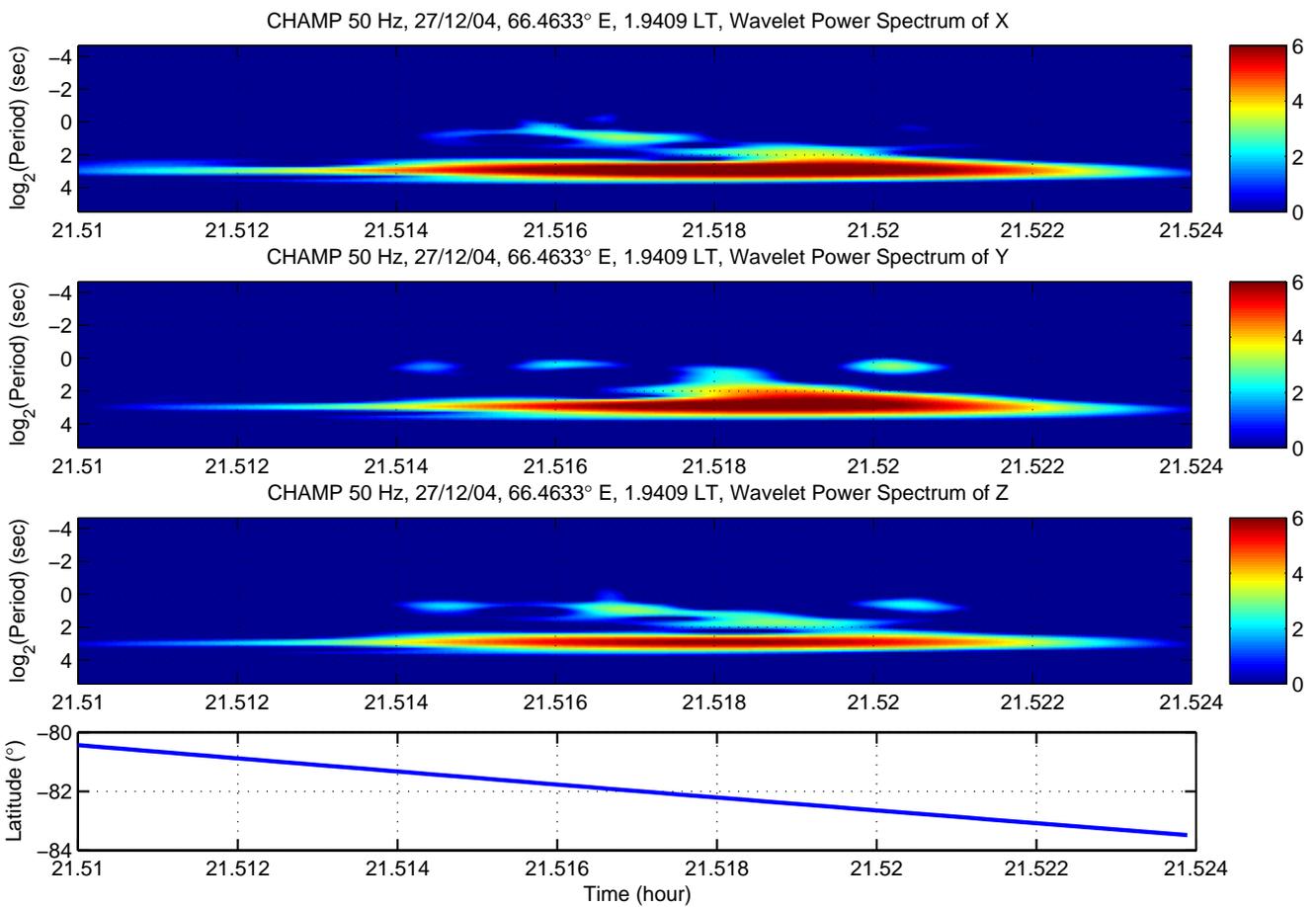}
\caption{
Zoom of Fig. 2, on the time interval 21.51--21.524 UT ($\simeq$ 50 s). This 
time span represents a relatively quiet magnetic period. The wavelet power 
spectra are dominated by a signal with a period centered around 7.49 s for X, 
7.46 s for Y and 7.5 s for Z component, all of which are close to the 7.56 s 
rotation period of magnetar SGR 1806-20. The symbols and scales are as in 
Fig. 2.}
\label{fig:3} 
\end{figure*}

\begin{figure*}
\noindent\includegraphics[scale=0.8]{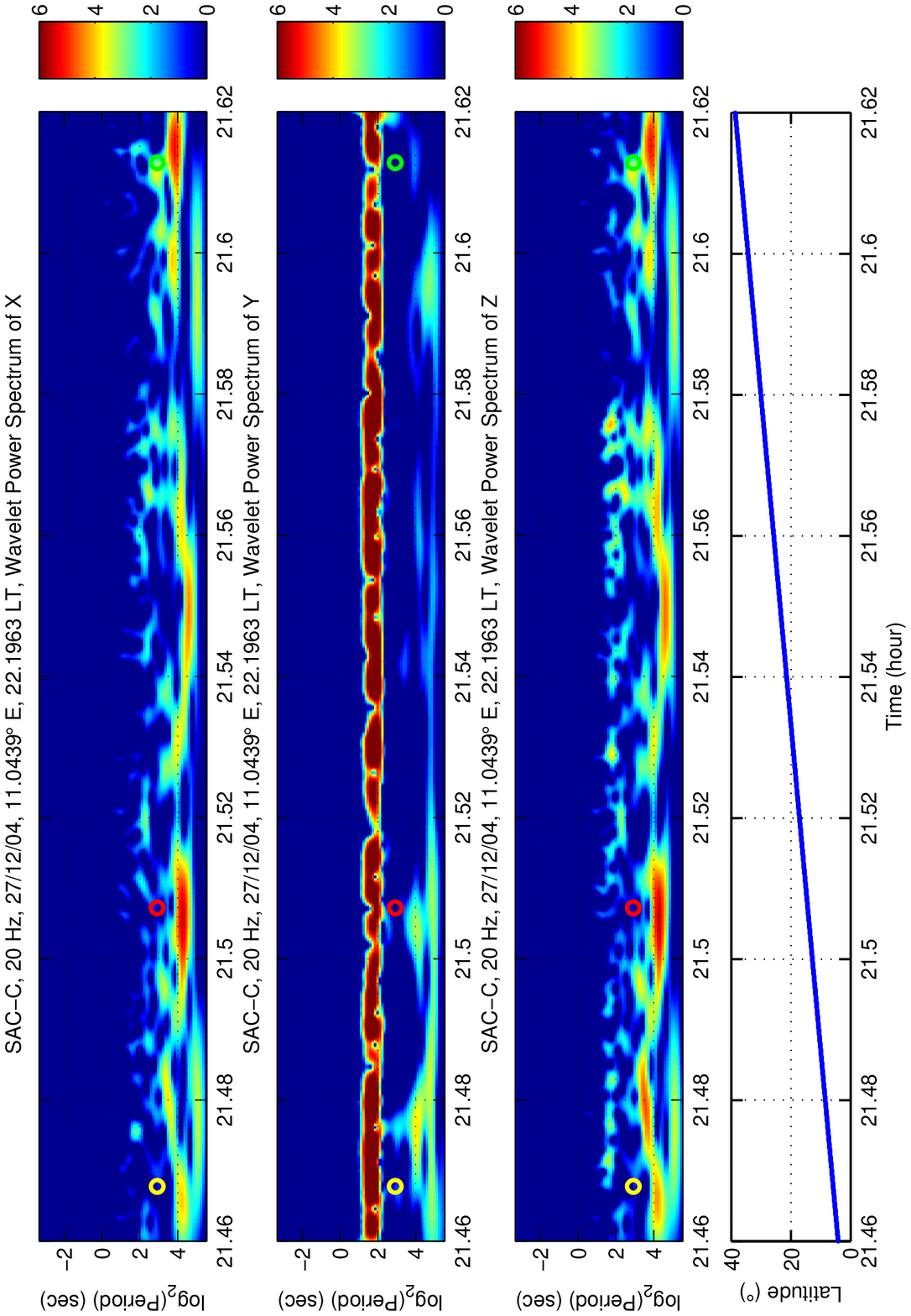}
\caption{
Wavelet analysis of the vector magnetic data provided by the SAC-C 
satellite on 27 December 2004. Diagrams as in Fig. 2.}
\label{fig:4} 
\end{figure*}

\begin{figure*}
\noindent\includegraphics[scale=0.8]{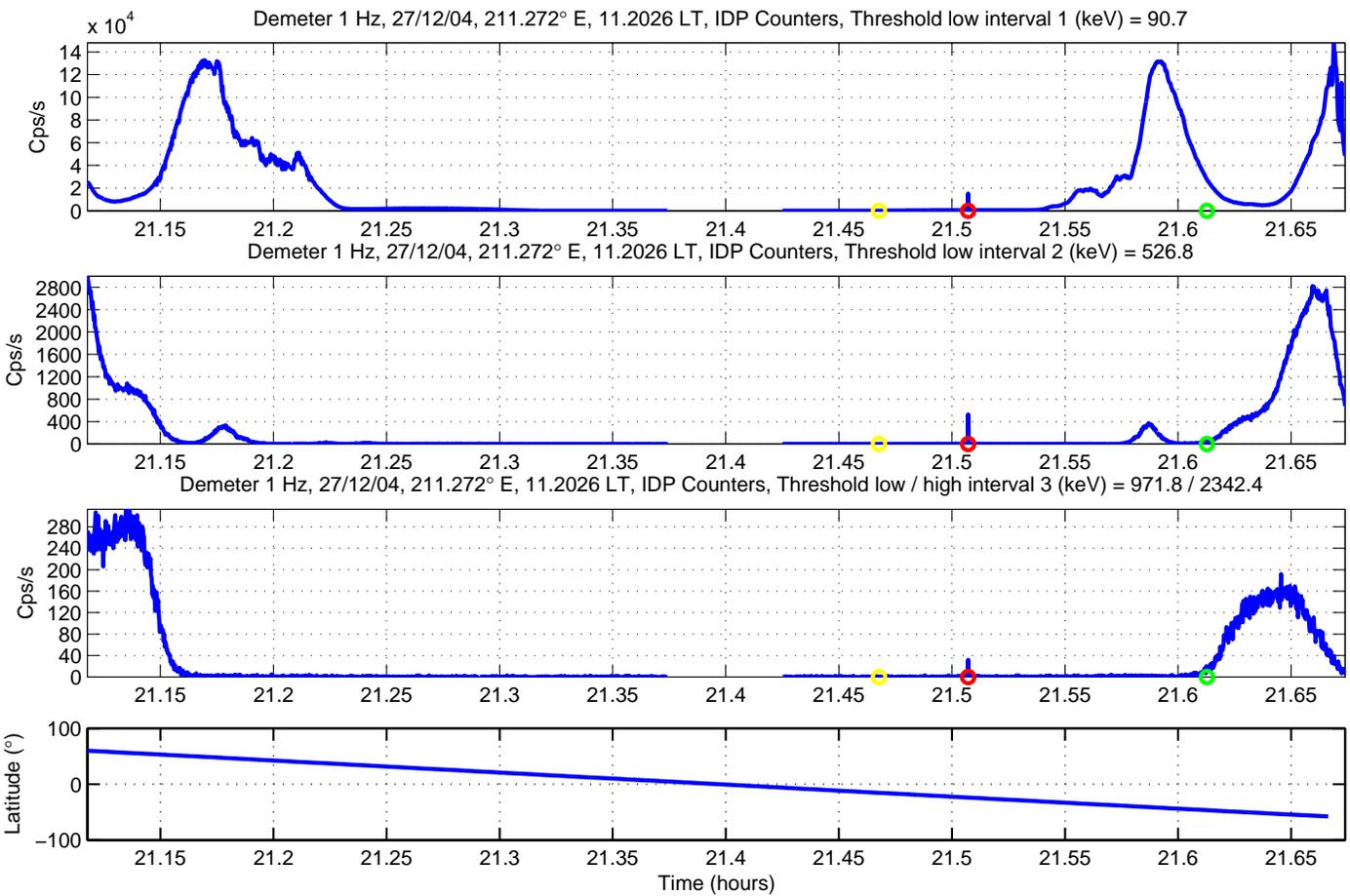}
\caption{
Energetic electron counter data from DEMETER satellite on 27 December 
2004 (orbit number: 02595\_0). From top to bottom, number of counts per 
second in 3 energy bands are shown. The 1 Hz data presented here were 
recorded in Survey mode. During the time interval 21.374 to 21.425 UT, 
DEMETER was operating in the Burst mode with a different sampling rate (see 
gap in the time series). The beginning of the flare from SGR 1806-20 is 
clearly observed as a peak in all three panels. The variation of satellite 
latitude with time is also given (c.f., lowermost panel).} 
\label{fig:5}
\end{figure*}

\end{document}